# Coincident Molecular Auxeticity and Negative Order Parameter in a Liquid Crystal Elastomer

D. Mistry[1], S. D. Connell[1], S. L. Micklethwaite[2], P. B. Morgan[3], J. H. Clamp[4] and H.F. Gleeson[1]

[1] School of Physics and Astronomy, University of Leeds, Leeds, LS2 9JT, UK
[2] Leeds Electron Microscopy And Spectroscopy centre, School of Chemical and Process Engineering, University of Leeds, Leeds, LS2 9JT, UK
[3] Eurolens Research, University of Manchester, Manchester, M13 9PL, UK
[4] UltraVision CLPL, Commerce Way, Leighton Buzzard, LU7 4RW, UK

"Auxetic" materials have the counter-intuitive property of expanding rather than contracting perpendicular to an applied stretch, formally they have negative Poisson's Ratios (PRs).[1,2] This results in properties such as enhanced energy absorption and indentation resistance, which means that auxetics have potential for applications in areas from aerospace to biomedical industries.[3,4] Existing synthetic auxetics are all created by carefully structuring porous geometries from positive PR materials. Crucially, their geometry causes the auxeticity.[3,4] The necessary porosity weakens the material compared to the bulk and the structure must be engineered, for example, by using resource-intensive additive manufacturing processes.[1,5] A longstanding goal for researchers has been the development of a synthetic material that has intrinsic auxetic behaviour. Such "molecular auxetics" would avoid porosity-weakening and their very existence implies chemical tuneability.[1,4–9] However molecular auxeticity has never previously been proven for a synthetic material.[6,7] Here we present a synthetic molecular auxetic based on a monodomain liquid crystal elastomer (LCE). When stressed perpendicular to the alignment direction, the LCE becomes auxetic at strains $\gtrsim 0.8$ with a minimum PR of -0.8. The critical strain for auxeticity coincides with the occurrence of a negative liquid crystal order parameter (LCOP). We show the auxeticity agrees with theoretical predictions derived from the Warner and Terentjev theory of LCEs.[10] This demonstration of a synthetic molecular auxetic represents the origin of a new approach to producing molecular auxetics with a range of physical properties and functional behaviours. Further, it demonstrates a novel feature of LCEs and a route for realisation of the molecular auxetic technologies that have been proposed over the years.

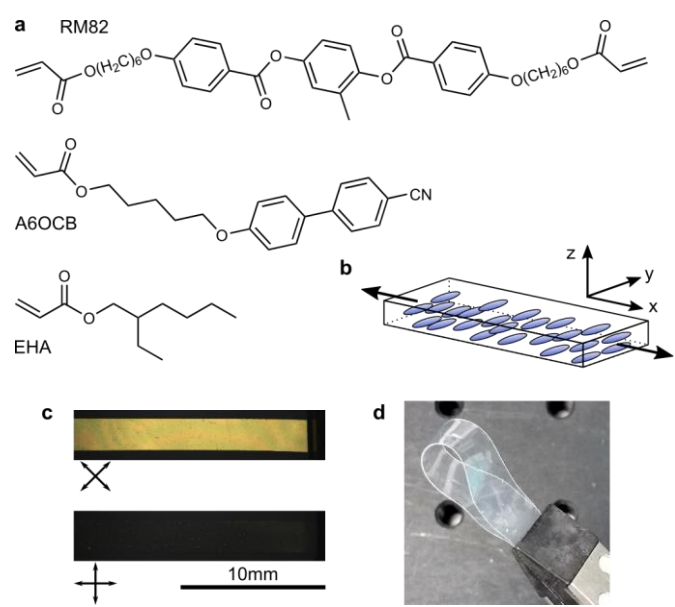

**Figure 1 | Composition and macroscopic appearance of LCE. a,** Acrylate monomers used to produce the auxetic side-chain LCE. **b,** Diagram of sample geometry with liquid crystal director aligned perpendicular to the film long axis. In this work strains are applied along the $x$ axis with the auxetic response observed along the $z$ axis. **c,** Polarising microscopy images of a unstrained LCE. **d,** Photograph of the final LCE showing its flexibility and high optical quality.

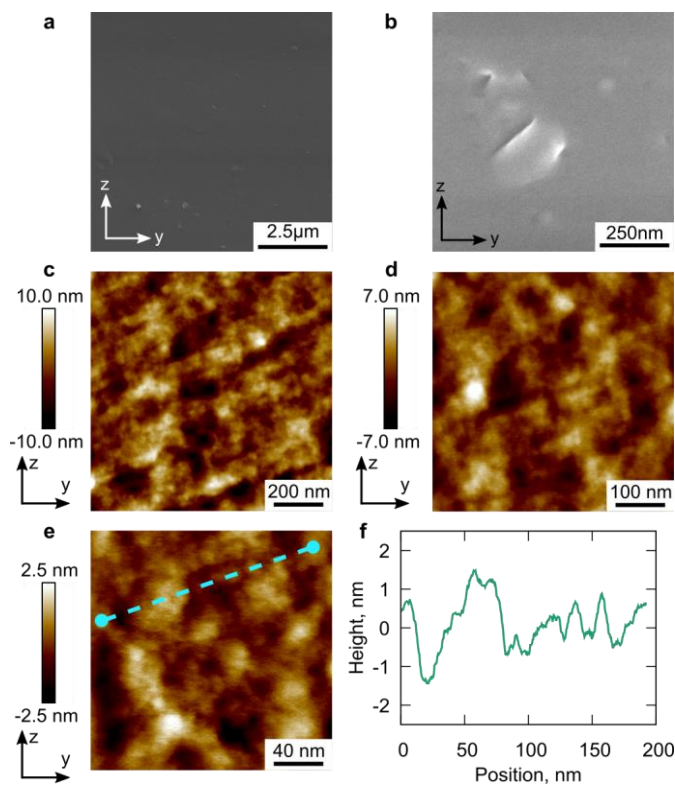

**Figure 2 | Micro- and nano-scopic structure of the LCE.** SEM micrographs (**a,b**) and AFM height maps (**c-e**) show the structures present from molecular to microscopic length scales. Axes used in **a-e** correspond to the coordinate set from Fig 1b. **f,** Profile of dashed line drawn across **e**.

The films of monodomain side-chain liquid crystal elastomer (LCE) were composed of 6-(4-cyano-biphenyl-4'-yloxy)hexyl acrylate (A6OCB, 34mol%), 2-ethylhexyl acrylate (EHA, 49mol%) and 1,4-bis-[4-(6-acryloyloxyhex-yloxy)benzoyloxy]-2-methylbenzene (RM82, 17mol%) (Fig. 1a) produced as described in Ref. 11 [11], further details are given in the methods and Supplementary Data Fig. 1. This LCE is a modified version of a commonly-used LCE which does not display auxetic behaviour.[12,13] In addition to introducing auxeticity, the modifications reduced the glass transition temperature of the LCE from 50°C to 14±1°C and ensured a room temperature nematic phase prior to polymerisation.[11] Monodomain LCEs were produced by polymerising the monomer mixture inside liquid crystal (LC) devices coated with a uniformly oriented planar alignment agent. The final LCE films had a birefringence $\Delta n = 0.08$ and thicknesses, $d$, in the range of 95-105μm. For mechanical characterisation, strips of dimensions ~2x18mm were cut with the nematic director at 89±1° to the film long axis (Fig. 1b).

The uniformity of monodomain alignment achieved in the LCE films is demonstrated by the high contrast and uniform polarising microscopy images shown in Fig. 1c, and by the high optical quality of the material shown in Fig. 1d. As explained above, the auxeticity of previously reported synthetic auxetics has been intimately tied to their porous geometry. By comparison, a molecular auxetic is expected to have zero material porosity across all length scales. Therefore it is important to here assess the nano- and micro-structure of the presented material. SEM images of a sample cross section exposed *via* freeze-fracturing show a homogeneous and largely featureless structure meaning the material has no porous geometry on length scales between tens of nanometers and micrometres. AFM, which is more sensitive to topographical features than SEM, shows a complex picture of the structure present on length scales from molecular to several hundreds of nanometers (Figs. 2c-f). From Fig. 2, two features are present: an extremely fine structure on the scale of a few nanometers; and a larger slow undulation on the scale of ~10 nm and with an amplitude of ~2 nm (comparable to the length of a molecule). Neither of these structures (seen most clearly from Figs. 2e and f) indicate the presence of material porosity on nanometer length scales. From Fig. 2 we conclude the LCE has zero detectable porosity.

**Table 1 | Testing parameters and PR at the maximum extension.**

| Test | Extension speed (% × $L_o$min$^{-1}$) | Temperature (±1°C) | $\epsilon_c$ - $\epsilon_x$ at emergence of auxeticity | PR at maximum extension |
|---|---|---|---|---|
| I | 16 | 28 | 0.99 | -0.34 |
| II | 7.5 | 28 | 0.78 | -0.51 |
| III | 1.0 | 24 | 0.80 | -0.52 |
| IV | 0.71 | 23 | 0.80 | -0.80 |

We studied the mechanical behaviour of the LCE films using a bespoke miniature tensile rig which simultaneously also allows observation of the polarising microscopy texture, thus providing an insight into the LCOP.[11] The films were extended in steps at various extension speeds, based on a percentage of the unstrained sample length, $L_o$, streched per minute, and at various temperatures as summarised in Table 1. Strains, $\epsilon_z$, (see Fig. 1b for the coordinate system used) were calculated *via* a constant volume assumption using strains, $\epsilon_x$ and $\epsilon_y$, measured in the $xy$ plane (see Methods). In each case $\epsilon_z$ approached a minimum before increasing (Table 1, Fig. 3a). Auxetic behaviour is demonstrated in the regions of further increasing $\epsilon_z$ where the sample is growing thicker in the direction transverse to the increasing applied strain, $\epsilon_x$. Fig. 3b shows for each test the strain-dependent Poisson's ratios, $\nu_{xz}$, calculated from the negative gradient of polynomials fitted to true strain versions of the curves shown in Fig. 3a. The curves confirm the emergence of auxetic behaviour above a critical strain, $\epsilon_c$, above which $\nu_{xz}$ becomes negative. $\epsilon_c$ shows a dependence on the sample temperature and on the speed at which it is extended (Table 1), most likely a result of the different conditions allowing different levels of stress relaxation between successive extensions. However, if the behaviour of $\nu_{xz}$ for each case is considered with respect to $\epsilon_c$, then it is seen that the magnitude of the auxetic response is in fact infact largely identical and hence independent of extension speed and temperature (Fig. 3c). In all cases **I-IV**, $\nu_{xz}$ monotonically decreases with strain. A minimum value of $\nu_{xz} = -0.8$ was recorded in test **IV** which was strained by the largest factor (~1.7) above $\epsilon_c$.

For test condition **I**, $\epsilon_z$ was measured in two ways: direct measurement, and from a constant volume assumption applied to measurements made in the $xy$ plane (see Methods, Fig. 3d). The excellent agreement between the two datasets is consistent with the LCE deforming at constant volume and hence at constant density. This result, which is to be expected for non-porous elastomers[10], validates the method of determining the strain $\epsilon_z$ *via* deformations in the $xy$ plane under a constant volume assumption and further, removes the possibility of the observations being caused by wrinkling effects.

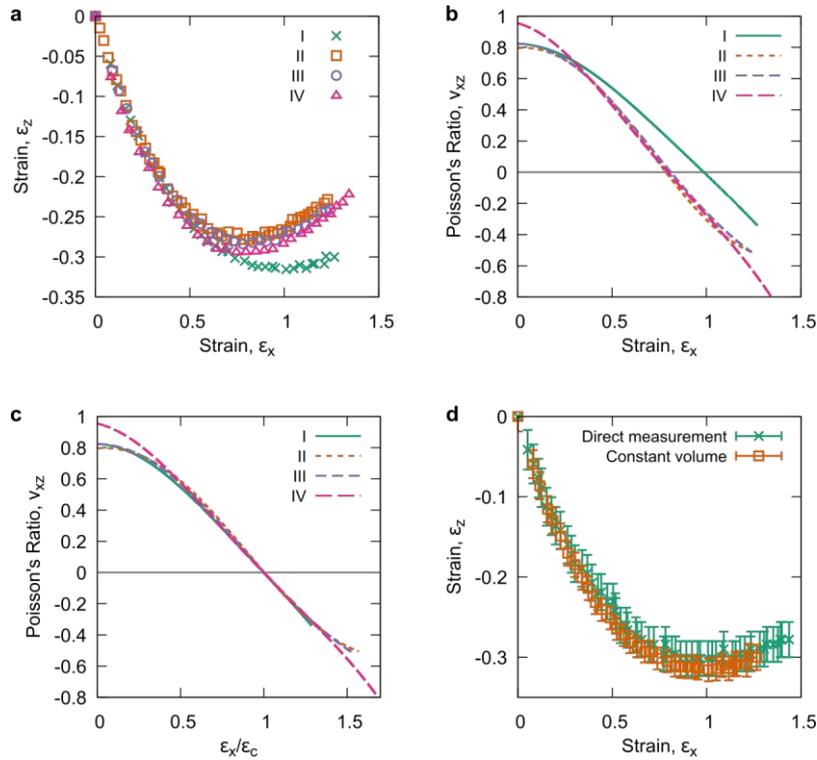

**Figure 3 | Measured strains, Poisson's ratio, and material volume conservation. a,** In all tests the strain $\epsilon_z$ initially decreases. Beyond the minimum the behaviour becomes auxetic. **b,** Strain-dependent Poisson's ratios extracted from (**a**) shows that $\nu_{xz}$ monotonically decreases and becomes negative. **c,** The Poisson ratios plotted relative to the strain corresponding to $\nu_{xz} = 0$ showing that the behaviour of $\nu_{xz}$ in all cases are largely identical. **d,** Consistency of density confirmed by the agreement between two strain datasets under conditions **I** collected *via* direct measurement and *via* a constant volume assumption applied to deformation measurements made in the $xy$ plane.

In all experiments, $\nu_{xy}$ was found to be positive at all strains tested (Supplementary Data Fig. 2). The anisotropy between $\nu_{xy}$ and $\nu_{xz}$ is to be expected for two reasons. Firstly, monodomain LCEs are well known to have inherent mechanical anisotropy.[10,14–16] Secondly, simultaneous volume conservation *and* Poisson's ratio isotropy is only possible in the specific case of $\nu_{xy} = \nu_{xz} = 0.5$, for which there is no auxetic behaviour.[17]

Fig. 4a shows polarising microscopy images of the LCE deformed under test conditions **I**. In the unstrained state the sample has an optical retardance $\Gamma = \Delta n \times d \approx 8000$ nm (~14[th] order). The progression to birefringence colours of increasing saturation indicates a decrease in retardance (to ~1[st]-3[rd] order between $\epsilon_x = 0.87$ and $\epsilon_x = 1.06$) that cannot be accounted for by the reducing sample thickness.[11] Such behaviour is associated with a reducing LCOP within the $xy$ plane. The black appearance at ($\epsilon_x = 1.14$) exists for all sample rotations with respect to the crossed polarisers (Supplementary Data Fig. 3) meaning there is zero birefringence and hence zero LC ordering *within the xy plane*. We have previously deduced that such an observation corresponds to a state of negative LCOP. Below we confirm this behaviour by applying the theory of LCEs pioneered by Warner and

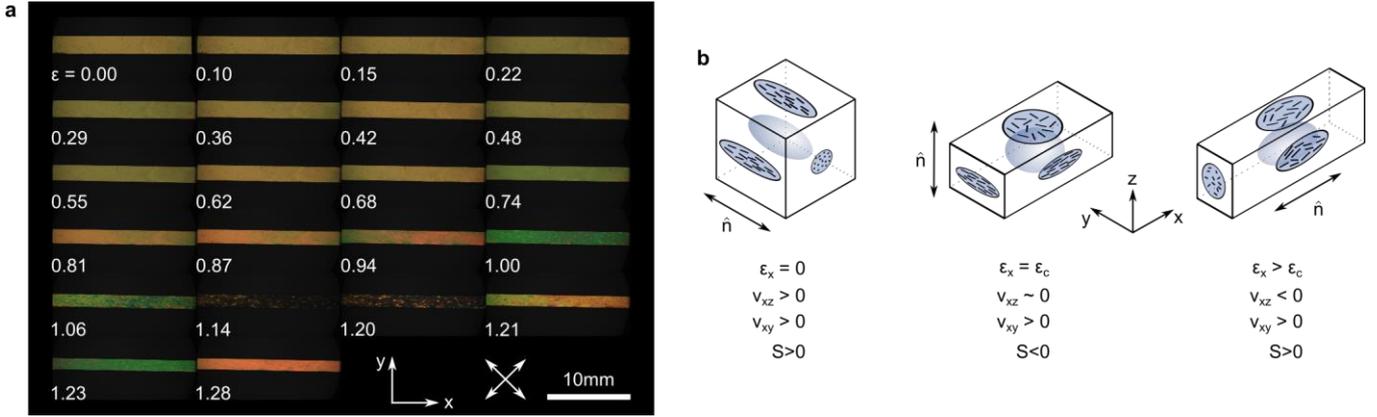

**Figure 4 | Relationships between LC order and the emergence of auxetic behaviour. a,** Polarising microscopy textures at each strain step of test **I**. The birefringence colours indicate the retardance initially decreases, becoming zero at $\epsilon_x = 1.14$, before increasing again. **b,** Model of the deformation described by: the relationships between the sample geometry (outline box), polymer conformation shape (enclosed ellipsoidal shapes) and LCOP (denoted as S) projected on each plane (rod arrangements). At the critical strain, $\epsilon_c$, the symmetry of the LC ordering corresponds to a negative order parameter with the director $\vec{n}$ lying parallel to the $z$ axis.

Terentjev (W&T).[11] Beyond $\epsilon_x \sim 1.20$, birefringence colours re-emerge and hence the in-plane LCOP increases, but with the director now parallel to the stress axis.[11]

The results presented here can be tested against the theory of W&T which speculated on the possibility of auxetic behaviour in aligned LCEs. This speculation has never previously been observed in experiments or simulations. The polymer chain conformation of a LCE is anisotropic and hence the step lengths of the Gaussian random walk are described by the effective step length tensor, $\underline{\underline{l}}$ which equals $\mathrm{Diag}(l_1, l_2, l_3)$ when appropriate axes are chosen.[10] Typically the effective step length tensor is uniaxial which, for instance, means $l_1 = l_\parallel$ and $l_2 = l_3 = l_\perp$ where the unique axis is aligned with the director. The step length anisotropy, $r = l_\parallel/l_\perp$, characterises the anisotropy of the polymer backbone and is fundamentally important to the physical phenomena observed in LCEs.[10] In their theory of LCEs, W&T speculated that an auxetic response of LCEs may be observed to begin at a deformation given by $(\lambda_x, \lambda_z) = (r_0^{1/3}, r_0^{-1/6})$, where $r_0$ is the step length anisotropy for the unstrained LCE and the $\lambda_i (= \epsilon_i + 1)$, are components of the deformation gradient tensor. Considering data for test **I** (chosen as we show corresponding polarising microscopy images in Fig. 4a) from Fig. 3a, we calculate using the critical strains of $(\epsilon_x, \epsilon_z) = (1.00 \pm 0.05, -0.31 \pm 0.02)$ values of $r_0 = 8.0 \pm 0.6$ and $r_0 9.3 \pm 1.6$ which are comfortably self-consistent. W&T theory also predicts the strain $\epsilon_z$ to return to zero by $\lambda_x = r_0^{1/2} \sim 9^{1/2} \sim 3$ and hence $\epsilon_x \sim 2$. While our samples all failed well before $\epsilon_x \sim 2$, inserting $\epsilon_x = 2$ into the 4$^{th}$ order polynomial fitted to the $\epsilon_x$-$\epsilon_z$ curve for test **I** (Fig. 3a) gives a strain $\epsilon_z = 0.077$, remarkably close to zero given the degree of extrapolation used.

Independent calculations of $r_0$ can be made by applying W&T theory to the strain at which we have observed zero birefringence (Fig. 4a). The coupling between the side-chain LC moieties and polymer backbone means the LCOP tensor and $\underline{\underline{l}}$ have a common symmetry. For the unstrained LCE, the director lies along the $y$ axis and so $\underline{\underline{l}} = \underline{\underline{l^0}} = \mathrm{Diag}(l_\perp^0, l_\parallel^0, l_\perp^0)$. At $\epsilon_x \approx 1.15$ the zero LCOP within the $xy$ plane means the effective step length tensor will be given by $\underline{\underline{l}} = \underline{\underline{l'}} = \mathrm{Diag}(l'_\perp, l'_\perp, l'_\parallel)$, which has principle axes parallel with those of $\underline{\underline{l^0}}$. Minimisation of the elastic free energy of a LCE, given by the "trace formula", provides predictions for the physical state of the LCE at $\epsilon_x \approx 1.15$. Ref 18[18] performs such a minimisation with respect to deformations along the $x$ and $y$ axes. From this, the following equations are derived for the deformation, $\lambda_y$, along the $y$ axis, and the elastic free energy, $F_{el}$, for a general deformed state of $\underline{\underline{l}}=\text{Diag}(l_x, l_y, l_z)$ which, like here, has principle axes parallel to those of $\underline{\underline{l^0}}$:

$$\lambda_y = \frac{1}{\sqrt{\lambda_x}}\left(\frac{l_y l_\perp^0}{l_z l_\parallel^0}\right) \text{ and } F_{el} = \frac{1}{2}\mu\left(\lambda_x^2 \frac{l_\parallel^0}{l_x} + \frac{2}{\lambda_x}\sqrt{\frac{l_\perp^0 l_\parallel^0}{l_y l_z}}\right), (1)$$

where $\lambda_x$ is the deformation along the extension direction.

By substituting $l_x = l_y = l'_\perp$, and $l_z = l'_\parallel$ into the above and minimising $F_{el}$ with respect to $\lambda_x$ we arrive at the following relationships:

$$\lambda_x^6 = \frac{r_0}{r'} \text{ and } \lambda_y^6 = \frac{1}{r' r_0^2}, (2)$$

where $r_0 = l_\parallel^0/l_\perp^0$ and $r' = l'_\parallel/l'_\perp$ are the step length anisotropies of the unstrained and strained states respectively. Note, the above does not placed any constraint on the value of $r'$ with respect to unity. From Fig. 4a we can deduce for test I that at the state of a negative LCOP $\lambda_x = 2.15 \pm 0.05$ and $\lambda_y = 0.67 \pm 0.05$. By Inserting these values in to equations (2) and solving, we find $r_0 = 10.2 \pm 1.6$ and $r' = 0.103 \pm 0.015$. This calculation of $r' < 1$ corresponds to the system adopting a uniaxial oblate polymer conformation. Previously we have determined values of $r_0 = 9.3$ and $r_0 = 3.8$ from opto-mechanical and thermal tests respectively.[11] At the time, comparisons of these values to those from LCEs of similar chemistries led us to conclude that the latter value was more likely to be accurate.[11] However, the several independent calculations of $r_0 \sim 9$ presented here now leads us to believe that this value is actually most likely to be correct. The self-consistency of values presented here demonstrates that W&T theory describes the physical behaviour of our material well and hence it can be used to make predictions on how the auxeticity of LCEs can be tuned. For instance the magnitude of the auxetic response could likely be increased by selecting an LCE with a greater value of $r_0$.

Using the calculated values of $r$ we can also extract values for the polymer chain backbone order parameter, $Q_b$, using the equation[18]

$$r = \frac{1+2Q_b}{1-Q_b} \therefore Q_b = \frac{r-1}{r+2} (3),$$

which gives $Q_b^0 = 0.74 \pm 0.03$ and $Q_b' = -0.41 \pm 0.01$ which, given the shared symmetry between the LCOP and $\underline{\underline{l}}$ tensors, are consistent with our present and previous deductions of the LCOP symmetry[11].

In Fig. 4b we bring together our results to illustrate the deduced relationships between: the macroscopic deformations; the polymer conformation shape; and the LC ordering with projections onto the $xy$, $yz$ and $zx$ planes. From this there appears to be a symmetry for the model of the system about the point of $v_{xz} = 0$ from which we speculate that the transition to an auxetic regime is required by symmetry.

The auxetic behaviour in the LCE we describe appears to be an inherent material property and not the consequence of a porous structure. Hence, this LCE is the first example of a synthetic molecular auxetic. This observation is particularly significant as LC polymers and LCEs have arguably been considered as the most promising type of synthetic material for displaying molecular auxeticity, but a negative Poisson's ratio has never previously been measured.[6,9,10,19–21] The maximum magnitude of negative PR observed in this system, $-0.8$, is larger than most values seen in naturally occurring molecular auxetics such as α-cristobalite ($\nu \sim -0.5$), and cubic metals (broad range of negative $\nu$ calculated between $0 \lesssim \nu \lesssim -0.8$).[22,23] From the close coincidence of $\epsilon_c$ with the negative LCOP state, and the consistency of values of $r$ calculated using W&T theory, we anticipate that the emergence of auxeticity is intrinsically linked to occurrence of a negative LCOP.

We finish by noting that by coupling auxeticity with other exciting properties of LCEs, such as soft-elasticity and shape-photoresponsivity, new prospects for LCE devices and applications are now apparent.[10,24–26] For instance, the use of azo-benzene LC groups offers possibilities in photo-patterning and photo-switching of the auxetic behaviour[26–29]. Moreover, as to the best of our knowledge a transparent auxetic material has never been previously observed, the presented LCE opens the future possibility of optical-auxetic sensors and devices.

## Methods

**Preparation of monodomain side-chain LCEs.** Materials were prepared as described in Ref. 11. Briefly, acrylate monomers: 6-(4-cyano-biphenyl-4'-yloxy)hexyl acrylate (A6OCB, 15mol%), 2-Ethylhexyl acrylate (EHA, 21mol%), 1,4-Bis-[4-(6-acryloyloxyhex-yloxy)benzoyloxy]-2-methylbenzene (RM82, 7mol%) (Fig. 1a) were mixed with the liquid crystal 4'- hexyloxybiphenyl (6OCB, 56mol%) and photoinitiator methyl benzoylformate (MBF, 1.6mol%). Use of 6OCB gave a broad nematic phase for alignment at room temperature. 6OCB and MBF were washed from the LCE following polymerisation. LC devices used for as moulds and for alignment of the monomer mixture were prepared according to the supplementary information of Ref. 11.

**SEM.** A section of the $yz$ plane (coordinate system shown in Fig. 1b) was exposed by freeze-fracturing a LCE sample. Prior to studying via SEM, the exposed faces were coated with a ~15nm conductive carbon layer using a Quorum Q150T E high vacuum evaporative carbon coater. A Hitatchi SU8230 SEM was used to image the exposed cross-sections from molecular to micron length scales. The single-pixel resolution in Fig. 2b is 1nm.

**AFM.** $yz$ cross-sectional samples (coordinate system shown in Fig. 1b) were prepared by encasing samples of LCE within a two-part epoxy glue and then freeze-fracturing by snapping off the exposed LCE. AFM images were acquired using a Bruker Dimension FastScan-Bio, using Bruker FastScan A probes, in air tapping mode at a frequency of 1.4 MHz. Images were acquired at a line rate of approximately 4 Hz at 1024 pixel resolution, then processed with a simple low order line flattening in Bruker Nanoscope Analysis v1.9.

**Mechanical testing.** The specification of the bespoke equipment used to measure the sample deformations and polarising microscopy textures is described at length in Ref. 11 and accompanying supplementary information. For most experiments the $xy$ plane of the sample was photographed at each strain step. From images taken at each strain step the strain $\epsilon_x$ was measured using the relative separation of features of the samples and the strain $\epsilon_y$ was measured from the changing sample width. Under the constant volume assumption, an initial volume element of dimensions $l_0 \times w_0 \times t_0$ will maintain a constant volume as the element is deformed to dimensions $l \times w \times t$. The strain, $\epsilon_z$, in the thickness direction can therefore be determined via

$$\epsilon_z = \frac{t}{t_0} - 1 = \frac{l_0}{l} \times \frac{w_0}{w} - 1.$$

In order to confirm the validity of the use of the constant volume assumption the thickness of the sample was directly measured by observing the deformation of the sample thickness directly. For this, the sample edge was illuminated via reflected light and the lens magnification was increased by a factor of 13. The camera resolution in this mode was 3μm. As the initial sample thicknesses was typically 100μm, the thickness could be measured with sufficient precision for confirmation of the constant volume assumption. From photographs taken, the thickness was determined by a python script which at each strain step automatically measured the thickness in each pixel column. The median value and median absolute deviation were taken as the sample thickness and error respectively, from which the strain $\epsilon_z$ was directly determined.

Strain-dependent Poisson's ratios were calculated by taking the negative gradient of fourth order polynomial functions fitted to $\log \epsilon_x$-$\log \epsilon_z$ and $\log \epsilon_x$-$\log \epsilon_y$ curves (*i.e.* true strain curves). The fitted polynomial functions were restricted to ensure they passed through the absolute data point for the unstrained sample (0,0).

**Sample optical retardance.** Prior to mechanical testing the optical retardance of samples was determined using a Leica DM2700 P polarising optical microscope equipped with a Berek compensator. A description of the optical principles of using a Berek compensator to measure optical retardance can be found in[30].

30. Ruiz, T. & Oldenbourg, R. Birefringence of tropomyosin crystals. *Biophys. J.* **54,** 17–24 (1988).
**Author contributions**
D.M., H.F.G J.C, and P.B.M conceived the research and designed the material. D.M. produced the material and performed the optical and mechanical characterisation. S.C. performed the AFM and analysed the results. S.L.M performed the SEM. D.M. analysed the optical, mechanical and opto-mechanical results. D.M. and H.F.G prepared the manuscript. All authors reviewed the manuscript.

**Acknowledgements**
D. Mistry thanks UltraVision CLPL and the EPSRC for a CASE PhD studentship and the Royal Commission for the Exhibition of 1851 for an Industrial Fellowship. The authors thank T Haynes and P Thornton for building equipment, Mark Warner for useful discussions and the Leeds Electron Microscopy And Spectroscopy centre (LEMAS).
**Author Information**
Correspondence and requests for materials should be addressed to Devesh Mistry

# Supplementary figures

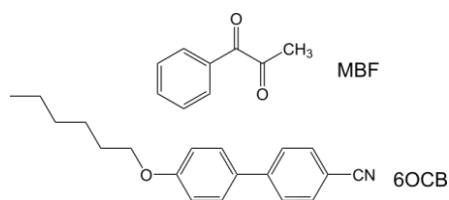

***Supplementary Fig. 1 | Additional chemicals used in preparation of LCEs.*** *Chemical components used in the monomer state and washed from the LCE following polymerisation. methyl benzoylformate (MBF) is a UV photoinitiator and 4'- hexyloxybiphenyl (6OCB) a liquid crystal required to give the monomer mixture the correct phase behaviour prior to polymerisation.*

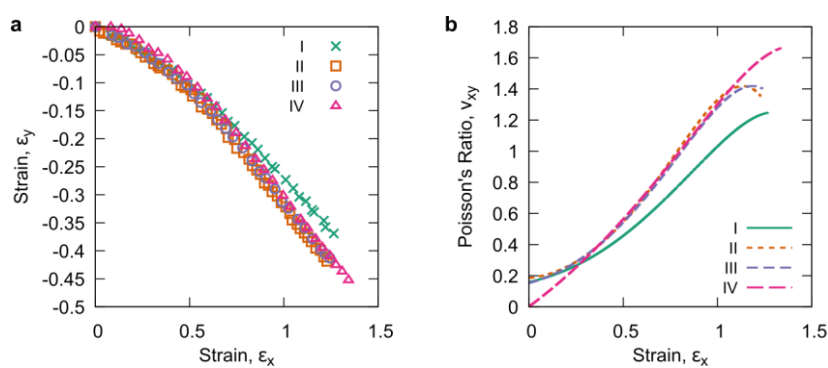

**Supplementary Data Fig. 2 | Strain and Poisson's ratio in the xy plane.** Measured strain $\epsilon_y$ (**a**) and Poisson's ratio $\nu_{xy}$ (**b**) relative to the imposed strain $\epsilon_x$ for each test **I** - **IV** detailed in Table 1

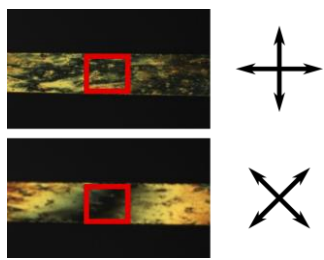

**Supplementary Data Fig. 3 | Zero LC order in $xy$ plane at the onset of auxeticity.** Polarising microscopy images taken at of a sample tested at parameters **II** at the point prior to the emergence of auxetic behaviour. The black appearance in the highlighted region in both photographs with the crossed polarisers indicates zero retardance and hence zero LC ordering within the image ($xy$) plane.